
\input harvmac

\def\aa{\alpha}
\def\b{\beta}

\def\d{\nabla}
\def\e{\epsilon_{\mu\nu\rho}}
\def\h{\hat}
\def\H{H_{\mu\nu\rho}}

\def\M{{\cal M}}
\def\ma{\pmatrix{ 1 & 0 \cr 0 & -1 \cr}}
\def\p{\varphi}
\def\pp{\partial}
\def\Q{{\cal Q}}
\def\t{\tilde}
\def\({\left (}
\def\){\right )}
\def\[{\left [}
\def\]{\right ]}
\gdef\journal#1, #2, #3, 19#4#5{
{\sl #1~}{\bf #2}, #3 (19#4#5)}

\lref\btz{M. Banados, C. Teitelboim, and J. Zanelli, ``Black Hole in
Three-Dimensional Spacetime", \journal Phys. Rev. Lett., 69, 1849, 1992;
M. Banados,  M. Henneaux, C. Teitelboim, and J. Zanelli, ``Geometry of
the 2+1 Black Hole", to appear in Phys. Rev. D}

\lref\bofw{ J. Balog, L. O'Raifeartaigh, P. Forgacs, and A. Wipf,
``Consistency of String Propagation on Curved Spacetimes: an $SU(1,1)$
Example", \journal Nucl. Phys., B325, 225, 1989; P. Petropoulos,
``Comments on $SU(1,1)$ String Theory", \journal Phys. Lett., B236, 151, 1990;
I. Bars and D. Nemeschansky, ``String Propagation in Backgrounds with Curved
Spacetime", \journal Nucl. Phys., B348, 89, 1991;
S. Hwang, ``No-Ghost Theorem for $SU(1,1)$ Strings", \journal Nucl. Phys.,
B354, 100, 1991.}
\lref\djh{S. Deser, R. Jackiw, and G. 't Hooft, ``Three Dimensional Einstein
Gravity: Dynamics of Flat Space", \journal Ann. Phys., 152, 220, 1984.}

\lref\dhvw{ L. Dixon, J. Harvey, C. Vafa, and E. Witten, ``Strings
on Orbifolds", \journal Nucl. Phys., B261, 678, 1985; \journal
Nucl. Phys., B274, 285, 1986.}

\lref\dvv{ R.~Dijkgraaf, E.~Verlinde and H.~Verlinde,
``String Propagation in a Black Hole Geometry,'' \journal Nucl. Phys., B371,
269, 1992.}

\lref\tseytlin{A. Tseytlin, ``Effective Action in Gauged WZW Models and
Exact String Solutions", Imperial/TP/92-93/10; ``Conformal Sigma Models
Corresponding to Gauged WZW
Theories", CERN-TH.6804/93, hep-th/9302083.}

\lref\garfinkle{D. Garfinkle, ``Black String Traveling Waves", \journal
Phys. Rev., D46, 4286, 1992.}
\lref\gershon{A. Hassan and A. Sen, ``Marginal Deformations of WZWN and Coset
Models from O(d,d) Transformations", preprint TIFR-TH-92-61, hep-th/9210121;
D. Gershon, ``Coset Models Obtained by Twisting
WZW Models...", Tel-Aviv preprint TAUP-2005-92; E. Kiritsis, ``Exact Duality
Symmetries in CFT and String Theory", CERN-TH.6797/93, hep-th/9302033.}

\lref\hape{S. Hawking and R. Penrose, ``The Singularities of Gravitational
Collapse and Cosmology", \journal Proc. Roy. Soc. Lond., A314, 529, 1970.}

\lref\witten{ E.~Witten, ``On String Theory and Black Holes,''
\journal Phys. Rev., D44, 314, 1991.}
\lref\hoho{ J.~Horne and G.~Horowitz,
``Exact Black String Solutions in Three Dimensions,''
\journal Nucl. Phys., B368, 444, 1992.}

\lref\buscher{ T. Buscher, ``Path Integral Derivation of Quantum
Duality in Nonlinear Sigma Models,''
\journal Phys. Lett., B201, 466, 1988 ;
``A Symmetry of the String Background Field Equations,''
\journal Phys. Lett., B194, 59, 1987.}

\lref\horowitz{G. Horowitz, ``The Dark Side of String Theory: Black Holes
and Black Strings", UCSBTH-92-32, hep-th/9210119.}

\lref\hhs{ J.~Horne, G.~Horowitz, and A.~Steif,
``An Equivalence Between Momentum and Charge in String Theory,''
\journal Phys. Rev. Lett., 68, 568, 1992.}
\lref\host{ G.~Horowitz and A.~Strominger, ``Black Strings
and $p$-Branes,''
\journal Nucl. Phys., B360, 197, 1991.}
\lref\rove{ M.~Ro\v cek and E.~Verlinde,
``Duality, Quotients, and Currents,'' \journal Nucl. Phys., B373, 630, 1992
hep-th/9110053.}
\lref\giro{A. Giveon and M. Ro\v cek, ``Generalized Duality in Curved
String Backgrounds", \journal Nucl. Phys., B380, 128, 1992.}

\lref\pete{ M. Perry and E. Teo, ``Non-Singularity of the Exact Two-Dimensional
String Black Hole", DAMTP R93/1, hep-th/9302037;
P. Yi, ``Nonsingular 2-D Black Holes and Classical String Backgrounds",
CALT-68-1852, hep-th/9302070.
This result was anticipated in \basf.}
\lref\basf{ I. Bars and K. Sfetsos, ``Conformally Exact Metric
and Dilaton in String
Theory on Curved Spacetime", \journal Phys. Rev., D46, 4510, 1992. }

\lref\sfetsos{ K. Sfetsos, ``Conformally Exact Results for $SL(2,R) \times
SO(1,1)^d/SO(1,1)$ Coset Models", USC-92/HEP-S1, hep-th/9206048, to appear
in Nucl. Phys.}

\lref\strominger{We thank A. Strominger for suggesting that a nonzero $\H$
might alter this result.}
\lref\basfn{I. Bars and K. Sfetsos, ``Exact Effective Action and
Spacetime Geometry in Gauged WZW Models", USC-93/HEP-B1, hep-th/9301047.}

\Title{\vbox{\baselineskip12pt\hbox{NSF-ITP-93-21}
\hbox{hep-th/9302126}}}
{\vbox{\centerline {Exact Three Dimensional Black Holes in String Theory}}}

\centerline{{Gary T. Horowitz}\footnote{$^*$}
{On leave from the Physics Department, University of California, Santa Barbara,
CA.}}
\vskip.1in
\centerline{\sl Institute for Theoretical Physics}
\centerline{\sl University of California}
\centerline{\sl Santa Barbara, CA 93106-9530}
\centerline{\sl gary@cosmic.physics.ucsb.edu}
\vskip .1in
\centerline{\sl Dean L. Welch}
\centerline{\sl Department of Physics}
\centerline{\sl University of California}
\centerline{\sl Santa Barbara, CA 93106-9530}
\centerline{\sl dean@cosmic.physics.ucsb.edu}

\bigskip
\centerline{\bf Abstract}

 A  black hole solution to three dimensional
general relativity with a  negative
cosmological constant has recently been found. We show that a slight
modification of this solution yields an
exact solution to string
theory. This black hole is  equivalent (under duality)
to the previously discussed
three dimensional black string solution.
Since the black string
is asymptotically flat and the black hole is asymptotically anti-de Sitter,
this suggests that strings are not affected by a negative cosmological
constant
in three dimensions.

\Date{2/93}


In a recent paper \btz,  Banados et. al. showed that there is a black hole
solution to three dimensional general relativity with a negative
cosmological constant. At first sight this is surprising, since the field
equation for this theory requires that, locally, the curvature is constant.
However they showed that by identifying certain points of three dimensional
anti-de Sitter space, one obtains a
solution with almost all of the usual features of a black hole.
In fact, there are a
two parameter family of inequivalent identifications leading to black holes
with mass $M$ and angular momentum $J$. Even though the curvature is constant,
the solutions have trapped surfaces, an event horizon, and nonzero Hawking
temperature. When $J\ne 0 $,
they also have
an ergosphere, and inner horizon. The solutions all approach
anti-de Sitter space (without identifications) asymptotically.

This solution is easily modified to obtain an exact solution to string theory.
One simply adds an antisymmetric tensor field $\H$ proportional to the
volume form $\e$. The reason is the following. There is a well known
construction (the Wess-Zumino-Witten model)
for obtaining a conformal field theory describing string
propagation on a Lie group. The natural metric on the
group $SL(2,R)$ is precisely the three dimensional anti-de Sitter metric.
So the WZW model based on $SL(2,R)$ is an exact conformal
field theory
describing string propagation on anti-de Sitter space \bofw.  The $\H$ field
is required by the Wess-Zumino term  and must be chosen so that the connection
with torsion $\H$ is flat. To obtain the black hole, one applies
the orbifold procedure \dhvw\ to obtain a conformal field theory describing
string propagation on the quotient space. This projects onto the states
which are invariant under the discrete group,
and adds the winding states.

This solution is of interest for several reasons. An exact four dimensional
black hole in string theory has not yet been found. A few years ago,
Witten showed \witten\
that an exact two dimensional black hole could be obtained
by gauging a one dimensional subgroup of $SL(2,R)$.
The three dimensional black hole has a number of advantages over the
two dimensional one. First, strings in three dimensions resemble higher
dimensional solutions in that there are an infinite
number of propagating modes. One can thus examine
their effect on Hawking evaporation. Second, the construction
is even simpler than the two dimensional black hole. One merely quotients by
a discrete subgroup rather than gauging a continuous one. In three dimensions,
there will  presumably be a tachyon, which can be removed by considering
the supersymmetric WZW model.

One of the most interesting properties of string theory is that different
spacetime geometries can correspond to equivalent classical solutions.
We will show that the three dimensional
black hole is equivalent to the
charged black string solution discussed earlier \hoho.
Some implications of this equivalence
for spacetime  singularities and the cosmological constant
problem will be considered.


We begin by reviewing the black hole solution discovered by Banados et. al.
\btz.
 Anti-de Sitter space can be
represented as the surface
\eqn\surface{ -x_0^2 -x_1^2 + x_2^2 +x_3^2 = - l^2 }
in the flat space of signature (-- -- + +)
\eqn\flatt{ ds^2 = -dx_0^2 -dx_1^2 + dx_2^2 + dx_3^2 }
It has curvature $R_{\mu\nu} = -(2/l^2) g_{\mu\nu}$.
This space is clearly invariant under $SO(2,2)$. The six independent
Killing vectors consist of two rotations (in the $(0,1)$ and $(2,3)$ planes)
and four boosts. A convenient way to parameterize the surface is to
choose two
commuting Killing vectors and let two of the coordinates be the parameters
along these symmetry directions. If $t$ and $\p$ are parameters along
the two rotations, the metric takes the familiar form
\eqn\ads{ds^2=-\( 1+{r^2\over l^2} \) dt^2 + \( 1+{r^2 \over l^2}\)^{-1} dr^2 +
     r^2 d\p^2}
To obtain the black hole, we let $\h t$ and $\h \p$ be parameters along
the boosts
in the (0,3) and (1,2) planes. Explicitly, set
\eqn\coord{\eqalign{ x_1 =&\h r \cosh \h \p   \qquad x_0 =
     \sqrt{l^2 -\h r^2}\cosh{(\h t/l)} \cr
    x_2 =&\h r\sinh \h\p \qquad x_3= \sqrt{l^2 -\h r^2}\sinh{(\h t/l)} }}
Notice that $\h r^2 >0$ only covers the region $x_1^2 - x_2^2 > 0$. Since this
is not the entire space, it might be more natural to use the radial coordinate
$\rho = \h r^2$ which takes both positive and negative values. For now, we will
continue to
use $\h r$ to maintain agreement with \btz. In terms of these coordinates,
anti-de Sitter space becomes
\eqn\newads{ds^2=\( 1-{\h r^2\over l^2} \) d\h t^2+
   \({\h r^2 \over l^2}-1 \)^{-1}d\h r^2
 + \h r^2 d\h \p^2}
Since $\h t$ and $\h \p$ are both parameters along a boost, they can take any
real value. If we identify $\h \p =\h \p + 2\pi$, \newads\
describes a black hole.
The surfaces of constant $\h t$ and $\h r<l$ are now compact trapped surfaces.
One could also identify $\h \p$ with a period other than $2\pi$. It turns out
that this corresponds to changing the mass of the black hole. This is
analogous to the fact that in the absence of
a cosmological constant,  the mass (in three dimensions)
is related to the deficit angle at infinity \djh. To add angular momentum, one
periodically identifies a linear combination of  $\h\p$ and $\h t$, rather than
$\h \p$ itself.

To be more explicit, choose two constants $r_+, r_-$ and introduce new
coordinates
$ \h t = (r_+ t/l) - r_- \p, \quad
\h \p = (r_+ \p/l) -( r_- t/l^2),\quad\h r^2=l^2 (r^2-r_-^2)/(r_+^2 - r_-^2)$.
Then the metric \newads\ becomes \btz
\eqn\bh{ ds^2 = \( M-{r^2\over l^2} \) dt^2 - J dt d\p + r^2 d\p^2 +
     \( {r^2 \over l^2} - M + { J^2 \over 4r^2} \)^{-1} dr^2 }
where the constants $M$ and $J$ are related to $r_\pm$ by
\eqn\rpm{ M= {r_+^2 + r_-^2 \over l^2} \qquad J={2r_+ r_- \over l}}
Identifying $ \p$ with $\p + 2\pi$, yields a two parameter
family of black holes.
By  paying careful attention to the surface terms in a Hamiltonian analysis
\btz, one finds that $M$ is  the mass and $J$ is the angular
momentum of the solution. In general, there are two horizons where
$\d_\mu r$ becomes null, which are located at $r = r_\pm$.
These two horizons coincide when $|J| = Ml$ which is the extremal limit.
The extremal black hole, as well as the massless solution $M=J=0$, cannot
be obtained by the above identifications. Instead, one must use null boosts.
The original anti-de Sitter space \ads\ is recovered when
$M=-1 $ and $J=0$.
The Killing vector $\pp /\pp t$
becomes null at $r^2 = Ml^2$ which lies outside the event horizon
$r=r_+$ when $J\ne 0$. This is similar to the ergosphere in the Kerr solution.
Physically, it means that an observer cannot remain at rest with respect
to infinity when she is close to the horizon.

What is the spacetime like near $r=0$? Since the curvature is constant, there
cannot be a curvature singularity. When $J=0$ and $M > 0$,  the $\p$
translation symmetry
has a fixed point in the  (1,2) plane. This causes the solution, near
$r=0$, to resemble the Taub-NUT solution and have incomplete null geodesics.
However, when $J\ne 0 $, the symmetry has no fixed points and the
spacetime is completely nonsingular. This is consistent with the singularity
theorems \hape\ even though the spacetime has trapped surfaces and satisfies
the strong energy condition, because there are closed timelike curves.
(Recall that the continuation past $r=0$ consists of $r^2 $ becoming
negative, so $\p$ becomes timelike.)
Banados et. al. argue that one should end the
spacetime at $r=0$, which avoids the causality problem but creates incomplete
geodesics.
However, this appears very unnatural.
The four dimensional Kerr
solution also has closed timelike curves inside the inner horizon. (Although
the vector $\pp /\pp \p$ is timelike only for $r <0$ in Kerr, there are closed
timelike curves through every point with $r < r_-$. It is likely that a similar
result holds here.) The closed timelike curves are not expected to produce
a physical violation of causality because of the instability of the inner
horizon.  String theory provides another reason for not ending
the spacetime at $r=0$. The WZW construction clearly
includes all regions of the spacetime, including $r^2<0$.

The Hawking temperature for this black hole is
\eqn\temp{ T= {r_+^2 - r_-^2 \over 2\pi r_+ l^2}}
(The factor of $l^2$ was omitted in \btz.)
These black holes do not evaporate completely in a finite time.
To see this, notice that since the temperature is reduced by rotation,
we can obtain a lower limit on the lifetime by setting $r_- = 0 $. Then
the temperature and the horizon size are both proportional to
$r_+ \sim \sqrt M$. In three dimensions, the energy flux in thermal radiation
is proportional to $T^3$, so $\dot M \propto M^2$ which implies $M(t) \propto
1/t$. Since their lifetime is infinite, small black holes will act like stable
remnants.


We now turn to string theory. We first consider these black
holes in the context of the low energy approximation,
and then consider the exact conformal field theory.
In three dimensions, the low energy string action is
\eqn\action{
S= \int d^3 x \sqrt{-g}\ e^{-2\phi} \[ {4\over k} +R+4(\nabla \phi)^2 -
 {1 \over 12} H_{\mu\nu\rho}
H^{\mu\nu\rho}\] }
The equations of motion which follow from this action are
\eqna\fdeq
$$\eqalignno{&R_{\mu\nu} + 2\d_\mu\d_\nu \phi
      - {1 \over 4} H_{\mu\lambda\sigma}
       {H_\nu}^{\lambda\sigma} =0   &\fdeq a \cr
 &\d^\mu (e^{-2\phi} H_{\mu\nu\rho}) = 0 & \fdeq b \cr
 &4\d^2\phi -4(\d\phi)^2 + {4\over k} + R -{1\over 12} H^2 = 0 &\fdeq c\cr}$$
A special property of three dimensions is that $\H$ must be proportional
to the volume form $\e$. If we assume $\phi = 0$, then \fdeq{b} yields
$\H   = (2/l)\e$   where $l$ is a constant with dimensions of length.
Substituting this form of $H$    into \fdeq{a} yields
\eqn\einstein{  R_{\mu\nu} = -{2\over l^2} g_{\mu\nu} }
which is exactly Einstein's equation with a negative cosmological constant.
The dilaton equation \fdeq{c} will also be satisfied provided
$k = l^2$. Thus every solution to three dimensional general relativity
with negative cosmological constant, is a solution to low energy
string theory with $\phi=0$, $\H = (2/l) \e$ and $k = l^2$. In
particular, the two parameter family of black holes \bh\ is a solution
with
\eqn\more{ B_{\p t} = {r^2\over l}, \qquad \phi = 0}
where $H=dB$. An earlier argument \horowitz\ claiming that three dimensional
black hole solutions to \fdeq{}\ do not exist assumed that $\H = 0$
\strominger.

We now consider the dual of this solution.
Duality is a well known symmetry of string theory that maps any solution
of the low energy  string equations \fdeq{}\ with a translational symmetry
to another solution. (Under certain conditions, the two solutions correspond
to equivalent conformal field theories \rove\giro.)
Given a solution $( g_{\mu\nu}  , B_{\mu\nu}, \phi)$ that is independent of
one coordinate, say $x$,
then ($\tilde g_{\mu\nu}, \tilde B_{\mu\nu},\tilde \phi$)
is also a solution where \buscher
\eqn\sigmadual{\eqalign{
 \tilde g_{xx} & = 1/g_{xx}, \qquad \quad \t g_{x\aa} = B_{x\aa}/ g_{xx} \cr
  \t g_{\aa\b} & = g_{\aa\b} - (g_{x\aa}g_{x\b} - B_{x\aa}B_{x\b})/g_{xx} \cr
   \t B_{x\aa} & = g_{x\aa}/g_{xx}, \qquad
	   \t B_{\aa\b} = B_{\aa\b} -2 g_{x[\aa} B_{\b]x}/g_{xx} \cr
     \tilde \phi & = \phi - {1\over 2} \ln g_{xx} \cr }}
 and $\aa,\b$ run over all directions except $x$.

Applying this transformation to the $\p$ translational symmetry of
the black hole solution \bh\more\ yields
\eqn\dual{\eqalign{ \widetilde{ds}^2 = \(M - {J^2 \over 4r^2}\) dt^2 +
{2\over l} dt d\p+&
{1\over r^2} d\p^2 + \({r^2 \over l^2} - M + { J^2 \over 4r^2}\)^{-1} dr^2 \cr
 \tilde B_{\p t} =-{J \over 2r^2} \qquad & \qquad \phi = -\ln r}}
To better understand this solution, we diagonalize the metric. Let
\eqn\newco{ t= {l(\h x- \h t ) \over (r_+^2 - r_-^2)^{1/2}},
  \qquad \p = {r_+^2\ \h t - r_-^2\ \h x \over (r_+^2 - r_-^2)^{1/2}},
    \qquad r^2 = l \h r}
Then the solution becomes
\eqn\bkst{\eqalign{\widetilde{ds}^2 = - \left(1-{\M\over \h r}\right) d\h t^2
	   +  \left(1-{\Q^2\over \M \h r}\right)d \h x^2
   + &\left(1-{\M\over \h r}\right)^{-1} \left(1-{\Q^2\over \M \h
r}\right)^{-1}
			      {l^2 \, d\h r^2 \over 4 \h r^2} \cr
\phi = -{1\over 2} \ln\h r l \>, \quad & \quad
B_{\h x \h t} = {\Q \over \h r}}}
where $\M = r_+^2 /l$ and $\Q = J/2$.
This is precisely the previously studied three dimensional charged
black string solution \hoho.
Notice that the charge of the black string is simply proportional to the
angular momentum of the black hole. The horizons of the black string are
at the same location as the black hole $r^2 = r_\pm^2$. Since $\p$ is
periodic, both $\h t $ and $\h x$ will in general be periodic. To avoid
closed timelike curves, one must go to the covering space.

Since the dual of the black hole is the black string, it must be possible
to dualize the black string and recover the black hole. This is a little
puzzling since it has
been shown \hhs\ that if you dualize \bkst\ on $\h x$,
you obtain a boosted uncharged
black string. The charge $\Q$ is dual to the momentum in the symmetry
direction $P_{\h x}$. However, one can apply duality to any translational
symmetry
$\pp /\pp \h x + \alpha \pp/ \pp \h t$.  If $\alpha < 1$, then
the dual is again a charged black string. If $\alpha = 1$ the result is
different. The Killing vector
$\pp /\pp \h x + \pp/ \pp \h t$
has norm $(\M^2 -\Q^2)/ \M \h r$, so it is spacelike
everywhere but  asymptotically  null. One can easily verify that
the dual of the black string \bkst\ with respect to this symmetry is precisely
the three dimensional black hole.

We now consider a few special cases. The dual of the nonrotating
black hole, $J=0$ and $M>0$, is the uncharged black string. This is
simply the two dimensional black hole cross $S^1$.
For the zero mass solution ($M=J=0$) and the extremal limit ($|J|=Ml$)
the dual is still given by \dual, but the transformation to the black string
breaks down. The duals are not the zero mass and extremal black string, but
rather these solutions superposed with a plane fronted wave. Setting $M=J=0$
in \dual, and introducing new coordinates $t=-v,\ \p = u l/2,\ r = e^{\h r/l}$
yields
\eqn\massless{ \widetilde{ds}^2 = -du dv + d\h r^2 +
  {l^2\over 4} e^{-2\h r /l} du^2}
This is a plane fronted wave in the presence of a dilaton \hhs.
Setting $J^2 = M^2 l^2$ in \dual, and introducing new
coordinates $t=-v/M,\ \p = l(v+ Mu)/2,\
r^2 = l\h r$, the metric becomes
\eqn\dualext{ \widetilde{ds}^2 = -\(1-{Ml \over 2\h r}\) du dv +
    {l^2 d\h r^2 \over 4\(\h r - {Ml \over 2} \)^2 }
    + {M^2 l \over 4\h r} du^2 }
This describes a wave of constant amplitude traveling along an extremal
black string \garfinkle.
Finally, recall that the
full anti-de Sitter space corresponds to $J=0$ and $M=-1$.
Inserting these values into \dual\ and setting $t= \h t + \p/l$ yields
\eqn\adsdual{ \widetilde{ds}^2 = -d\h t^2 + \(1+ { r^2 \over l^2} \)^{-1} dr^2
    + \( {r^2 + l^2
    \over r^2 l^2} \) d \p^2}
which is the product of time and the dual of the two dimensional Euclidean
black hole.

We now turn to the exact conformal field theory description of the black hole.
As mentioned earlier, this is in terms of the $SL(2,R)$ WZW model.
One can parameterize $SL(2,R)$ by
\eqn\defgp{ g= \pmatrix { x_2 + x_1 & x_0+x_3 \cr
	    x_0 - x_3 & x_2 - x_1 \cr}}
The statement that this matrix has unit determinant
is just equation \surface\ with $-l^2$ replaced by $1$. The difference in sign
causes the
induced metric to have signature (+ -- --). One can correct for this
and reintroduce the scale
by multiplying the WZW action by $-k$. In other words, one considers the level
$k$ WZW model.  For the noncompact group $SL(2,R)$,
$k$ is not required to be an integer. The central charge is
$ c = {3k / (k-2)} $, so
$c=26$ when $k=52/23$.  One can also consider larger values of $k$ and
take the product of this black hole with an internal conformal field theory.
In terms of the group, translations of $\h\p$ in
\newads\ correspond to the axial symmetry
\eqn\axial{ \delta g = \epsilon\[ \ma g + g \ma \] }
while translations of $\h t$ correspond to the vector symmetry
\eqn\vector { \delta g = \epsilon\[ \ma g - g \ma \]}
The general black hole is obtained by quotienting under a discrete
subgroup of a linear combination of these symmetries. This can be carried
out by the standard orbifold construction \dhvw.

There is a close connection between the two and three dimensional black holes
in string theory.  Witten has shown \witten\
that the two dimensional black hole can
be obtained by starting with the $SL(2,R)$ WZW model and gauging the axial
symmetry \axial. If one gauges the vector symmetry,
one obtains the dual of the black hole, which turns out to have
the same geometry.
One cannot gauge a general linear combination of the symmetries
because of an anomaly.

Rocek and Verlinde have shown \rove\ that for a positive definite target
space having a spacelike symmetry with compact
orbits, the low energy duality \sigmadual\
corresponds to an equivalence between exact
conformal field theories. For Lorentzian target spaces, equivalence has not
yet been rigorously established. In addition to the obvious difficulty of
convergence of the functional integral, there are other issues involving
potential closed timelike curves. Nevertheless, since the equivalence
does not explicitly depend on the signature of the target space, one expects
it to hold in this case also. The effect of duality on a WZW model has been
investigated \rove\gershon. If one
dualizes with respect to an axial or vector
symmetry, the result (after a simple
shift of coordinate) is just the product of $U(1)$ and the WZW model with this
symmetry gauged \rove. This explains why the dual of the nonrotating
black hole is just the product of the two dimensional black hole and $U(1)$.
It also explains why the dual of anti-de Sitter is  the product of time
and the dual of the Euclidean black hole. However, we have seen that
the dual of the general rotating
black hole is  the charged black string which is not a simple product.
The exact conformal field theory associated with this solution is also
known \hoho.
One starts with the group $SL(2,R) \times U(1)$ and gauges the axial
symmetry \axial\ of $SL(2,R)$ together with rotations of $U(1)$.

Solutions to the low energy field equations
cannot be trusted in regions of large curvature. But since the three
dimensional
black hole has constant curvature  (which is small for large $k$),
it should be a good
approximation everywhere. In fact, for a WZW model, the
exact metric differs from the low energy approximation only by an overall
rescaling \tseytlin\basfn. So the exact three dimensional
black hole metric is simply
proportional to \bh,
and (for $J \ne 0 $) is nonsingular. (Although the stability of the inner
horizon and the effect of the closed timelike curves remain to be
investigated.)
A candidate for the exact two dimensional black hole metric has been found
\dvv\basf\tseytlin.
It has recently been shown \pete\
that this metric is also free of curvature singularities
(although the dilaton diverges inside
the event horizon). This can be viewed as increasing evidence that
exact black holes in string theory do not have curvature singularities.
But the evidence is far from conclusive.

A candidate for the exact three dimensional black string metric has
also been found \sfetsos\basfn, which does have a curvature singularity.
However, the fact that the solution is equivalent
to one without a curvature  singularity, suggests
that it may also correspond to a nonsingular conformal field theory.

Perhaps the most remarkable consequence of the equivalence between the
black hole and black string comes from the fact that the black hole
is asymptotically anti-de Sitter while the black string is asymptotically
flat. Since they are equivalent, it suggests that a negative cosmological
constant has no effect on strings in three dimensions. The reason that
the asymptotic structure of the spacetime changes under duality
is that the length of the circles parameterized by $\p$ does
not approach a constant at infinity. This phenomenon has been noticed
before, but in previous examples the asymptotic behavior of the dual space
did not have a simple physical interpretation. For example, consider
four dimensional Minkowski spacetime $ds^2 = - dt^2 + dr^2 + r^2 ( d\theta^2
+ \sin^2 \theta d\p^2)$. If we dualize on $\p$ the metric is identical except
that $g_{\p\p}$ is changed to $(r^2\sin^2\theta)^{-1}$ which is singular along
the axis $\theta = 0, \pi$. The three dimensional
black hole seems to be the first example
in which two different asymptotic behaviors each have a simple physical
interpretation.

This suggests a novel resolution of the cosmological constant problem.
Perhaps a solution with a cosmological constant in string theory
is equivalent to one without. Strings may not be affected by a cosmological
constant. Unfortunately, a straightforward generalization of our results
to higher dimensions, does not relate a solution with a cosmological constant
to one without.  One can start with the charged black
string in $D$ dimensions \host\ which is asymptotically flat,
and dualize with respect to a symmetry that is spacelike but asymptotically
null. The
result is a metric which is neither asymptotically flat nor asymptotically
anti-de Sitter. However, given our elementary
understanding of duality in string theory, one cannot rule out the possibility
that this effect will play a role in resolving the cosmological constant
problem.

\vskip 1cm

\centerline{Acknowledgments}
It is a pleasure to thank T. Banks, A. Giveon, J. Horne,
M. Rocek, and A. Strominger for discussions.
This work was supported in part by NSF Grants  PHY-8904035 and
PHY-9008502.

\listrefs

\end